\documentclass[%
twocolumn,
superscriptaddress,
 amsmath,amssymb,
 aps,
pra,
floatfix
]{revtex4-2}

\usepackage{color}

\usepackage{subfigure}

\usepackage{graphicx}
\usepackage{bm}
\usepackage{amsmath}
\usepackage{amssymb}
\usepackage{braket}
\usepackage{hyperref}
\usepackage{xcolor}

\begin{document}
\setcounter{page}{1}



\title{Finite-size security of QKD: 
comparison of three proof techniques}

\author{Gabriele Staffieri}
\affiliation{Dipartimento Interateneo di Fisica, Università di Bari, 70126 Bari, Italy}
\affiliation{INFN, Sezione di Bari, 70126 Bari, Italy}

\author{Giovanni Scala}
\affiliation{Dipartimento Interateneo di Fisica, Politecnico di Bari, 70126 Bari, Italy}
\affiliation{INFN, Sezione di Bari, 70126 Bari, Italy}

\author{Cosmo Lupo}
\affiliation{Dipartimento Interateneo di Fisica, Università di Bari, 70126 Bari, Italy}
\affiliation{Dipartimento Interateneo di Fisica, Politecnico di Bari, 70126 Bari, Italy}
\affiliation{INFN, Sezione di Bari, 70126 Bari, Italy}

\begin{abstract}
\noindent
We compare three proof techniques for composable finite-size security of quantum key distribution under collective attacks, with emphasis on how the resulting secret-key rates behave at practically relevant block lengths.
As a benchmark, we consider the BB84 protocol and evaluate finite-size key-rate estimates obtained from entropic uncertainty relations (EUR), from the asymptotic equipartition property (AEP), and from a direct finite-block analysis based on the conditional min-entropy, which we refer to as the finite-size min-entropy (FME) approach.
For BB84 we show that the EUR-based bound provides the most favorable performance across the considered parameter range, while the AEP bound is asymptotically tight but can become overly pessimistic at moderate and small block sizes, where it may fail to certify a positive key.
The FME approach remains effective in this small-block regime, yielding nonzero rates in situations where the AEP estimate vanishes, although it is not asymptotically optimal for BB84.
These results motivate the use of FME-type analyses for continuous-variable protocols in settings where tight EUR-based bounds are unavailable, notably for coherent-state schemes where current finite-size analyses typically rely on AEP-style corrections.
\end{abstract}



\maketitle

\section{Introduction}
\noindent
Classical public-key cryptography derives its practical security from the presumed computational hardness of problems such as integer factorization and discrete logarithms~\cite{rivest1978method}.
This paradigm is vulnerable to algorithmic and technological advances.
In particular, large-scale fault-tolerant quantum computers would undermine widely deployed public-key infrastructures through Shor's algorithms for factoring and related problems~\cite{shor1994algorithms,shor1999polynomial}.
This prospect has stimulated intense activity on quantum-resistant approaches to secure communication.

Quantum key distribution (QKD) offers an information-theoretic route to key establishment whose security is rooted in the laws of quantum mechanics rather than in computational assumptions~\cite{pirandola2020advances,Gisin2002,PortmannRenner2022RMP,Pirandola2020,Zapatero2023,Ghoreishi2025a}.
In QKD, two distant parties (Alice and Bob) exchange quantum signals and use authenticated public communication to distill a shared secret key.
The security mechanism ultimately traces back to the information--disturbance trade-off: any attempt to gain information about nonorthogonal quantum states necessarily induces disturbances that can be detected in the observed statistics.
Once established, the key can be used with symmetric cryptographic primitives (e.g.~the one-time pad or modern authenticated encryption) to protect data.

QKD protocols are commonly classified as discrete-variable (DV) or continuous-variable (CV) depending on the encoding and measurement strategy.
In DV QKD, information is encoded in finite-dimensional degrees of freedom, such as polarization or time-bin qubits, and recovered with single-photon detection~\cite{Gisin2002,pirandola2020advances}.
In CV QKD, information is encoded in field quadratures and measured with coherent detection (homodyne or heterodyne), enabling high-rate implementations with standard telecom components~\cite{pirandola2020advances,Pirandola2020}.
A central theoretical and practical issue in both settings is the finite-size regime: real systems exchange a finite number of signals, so parameter estimation is affected by statistical fluctuations and the achievable key length must be certified with composable security guarantees.
This regime is particularly relevant in CV QKD, where large data blocks are often required to overcome finite-size penalties and extract a positive key~\cite{jain2022practical}.

A variety of proof techniques have been developed to establish QKD security in composable frameworks~\cite{renner2008security,renner2004universallycomposableprivacyamplification,PhysRevA.72.012332,PhysRevLett.95.080501,PhysRevLett.100.200501,tomamichel2012framework,Devetak_2005}.
Among the most widely used approaches in DV QKD are proofs based on entropic uncertainty relations (EUR), which directly connect observed error statistics to bounds on the adversary's information~\cite{PhysRevLett.60.1103,krishna2001entropicuncertaintyprinciplequantum,PhysRevLett.103.020402,tomamichel2011uncertainty}.
In many practical finite-size analyses, one also encounters bounds derived from the asymptotic equipartition property (AEP), which relates smooth min-entropies to von Neumann entropies with finite-size corrections that typically scale as $1/\sqrt{N}$~\cite{cover1999elements,5319753,tomamichel2012framework}.
A less explored alternative is to bound the conditional min-entropy more directly at finite block length, without passing through an AEP-type approximation; in this work we refer to this strategy as a finite-size min-entropy (FME) approach and we build on the fidelity-based characterization of guessing probabilities~\cite{bratzik2011min,bunandar2020numerical,staffieri2025finite}.

In this paper we provide a focused comparison of three proof strategies---EUR, AEP, and FME---in the finite-size regime under the assumption of collective attacks.
As a case study we consider the BB84 protocol~\cite{bennett2014quantum}, which remains a canonical benchmark for DV QKD and a useful reference point for assessing the tightness of finite-size bounds.
We find that EUR-based bounds yield the strongest key-rate estimates in our analysis.
The AEP approach is asymptotically optimal but may become too conservative at moderate block sizes, where it can fail to certify a positive key.
By contrast, the FME approach can remain effective at smaller block sizes where AEP yields a vanishing rate, although it does not provide an asymptotically tight bound for BB84.
These observations suggest that FME-type tools may be particularly valuable for CV protocols in scenarios where tight EUR-based bounds are not available, such as coherent-state schemes for which standard security proofs often rely on AEP-style estimates in the collective-attack model.

The remainder of the paper is organized as follows.
We introduce the BB84 case study in an entanglement-based representation and specify the channel parameters accessible to Alice and Bob.
We then derive the finite-size min-entropy bound in terms of an adversarial guessing probability and compare the resulting finite-size key-rate estimates with those obtained from AEP and EUR techniques.
Finally, we present a numerical comparison of the three approaches and discuss the implications for finite-size security analyses beyond DV QKD.

\section{Case study: BB84}

We use the BB84 protocol~\cite{bennett2014quantum} as a representative DV QKD scheme and adopt its entanglement-based (EB) formulation, which is convenient for security analysis.
In each round Alice prepares a maximally entangled pair,
\begin{equation}
|\Phi^+\rangle_{AA'}=\frac{|00\rangle_{AA'}+|11\rangle_{AA'}}{\sqrt{2}}
=\frac{|\texttt{++}\rangle_{AA'}+|\texttt{--}\rangle_{AA'}}{\sqrt{2}},
\end{equation}
where $|\pm\rangle = (|0\rangle+|1\rangle)/\sqrt{2}$,
keeps system $A$, and sends system $A'$ to Bob through an insecure quantum channel, where it is received as system $B$.
For a block of $N$ rounds, we assume independent and identically distributed (i.i.d.) uses of the channel, corresponding to the collective-attack model.

A general collective attack can be represented by a Stinespring dilation: there exists an environment $E$, an initial state $|\phi\rangle_E$, and a unitary $U_{[A'\to B]E}$ such that
\begin{align}
|\Phi^+\rangle_{AA'}&\otimes|\phi\rangle_E
\ \longmapsto\nonumber\\
&(\mathbb{I}_A\otimes U_{[A'\to B]E})\,
|\Phi^+\rangle_{AA'}\otimes|\phi\rangle_E
=:|\Theta\rangle_{ABE}.
\end{align}
The bipartite state shared by Alice and Bob is then
$\rho_{AB}=\mathrm{Tr}_E\!\left(|\Theta\rangle\langle\Theta|\right)$,
while Eve is granted access to $E$.
From Alice and Bob's perspective $\rho_{AB}$ is unknown, but it is constrained by symmetries of the preparation stage and by parameter estimation.

Alice and Bob perform local measurements in the $\mathbb{Z}$ basis, $\mathbb{Z}=\{|0\rangle,|1\rangle\}$, or in the $\mathbb{X}$ basis, $\mathbb{X}=\{|\texttt{+}\rangle,|\texttt{-}\rangle\}$, and they keep only the rounds in which their basis choices match (sifting).
In the EB model, the reduced state on Alice's side is maximally mixed,
\begin{equation}
\rho_A=\mathrm{Tr}_B(\rho_{AB})=\frac{\mathbb{I}_A}{2}.
\label{rhoAmixed}
\end{equation}
They also estimate the quantum bit error rate (QBER), defined as the probability that their outcomes disagree when measuring in the same basis:
\begin{align}
\mathrm{QBER}^{(\mathbb{Z})}
&=\mathrm{Tr}\!\left[\left(|01\rangle\langle01|+|10\rangle\langle10|\right)\rho_{AB}\right], \nonumber\\
\mathrm{QBER}^{(\mathbb{X})}
&=\mathrm{Tr}\!\left[\left(|\texttt{+-}\rangle\langle\texttt{+-}|+|\texttt{-+}\rangle\langle\texttt{-+}|\right)\rho_{AB}\right].
\label{subb1}
\end{align}
These observed parameters restrict the set of admissible states $\rho_{AB}$ compatible with the experiment and are the only inputs needed by the security bounds compared in this work.

In addition to errors, optical loss reduces the number of detected signals.
We model loss through the channel transmittance $\eta\in[0,1]$, which depends on the communication distance $d$ and the fiber attenuation coefficient $a$ (in ${\rm dB}/{\rm km}$) according to
\begin{equation}
\eta = 10^{-\frac{1}{10}ad}.
\label{transm}
\end{equation}
After the quantum communication stage, Alice and Bob apply classical post-processing steps: parameter estimation (to bound the QBER), information reconciliation (to correct discrepancies between their raw keys), and privacy amplification (to extract a shorter key that is close to uniform and independent of Eve).
Throughout, we assume reverse reconciliation, meaning that Alice corrects her data to match Bob's.
We denote by $\mathrm{leak_{EC}}$ the number of bits revealed during error correction over the authenticated public channel.

\section{Finite-size security from the min-Entropy}
\label{minent}

The aim of this section is to lower bound the number of secret bits $\ell$ that Alice and Bob can extract after exchanging a finite block of signals.
In reverse reconciliation, the relevant quantity is the conditional min-entropy of Bob's raw key given Eve's side information.
For definiteness we take Bob's key-generating measurement to be in the $\mathbb{Z}$ basis and denote its classical outcome by $Z$.

According to the leftover hash lemma~\cite{tomamichel2011leftover}, if $n$ signals are used for key extraction under collective attacks, then privacy amplification can produce a secret key of length
\begin{equation}
\ell \ge n\,H_{\min}(Z|E)
+2\log_2\!\bigl(\sqrt{2}\,\epsilon_{\rm h}\bigr),
\end{equation}
where the secret key rate reads as $r=(\ell-\ell_{\mathrm{leak}})/N$, with  $\ell_{\mathrm{leak}}$ accounts for information revealed during error correction and $\epsilon_{\rm h}$ quantifies the failure probability of the randomness extraction step.
The overall composable security parameter also includes the (nonzero) error-correction failure probability $\epsilon_{\rm EC}$, as specified later when we combine all failure events.

Within the FME approach, we evaluate $H_{\min}(Z|E)$ through its operational relation to Eve's optimal guessing probability.
Following~\cite{coles2012unification}, one can write
\begin{equation}
H_{\min}(Z|E) = -\log_2 P_{\rm g}(Z|E),
\label{Coles00}
\end{equation}
where $P_{\rm g}(Z|E)$ is the maximum probability that an adversary, by measuring $E$, correctly guesses the outcome of Bob's $\mathbb{Z}$ measurement.
Moreover, $P_{\rm g}(Z|E)$ can be expressed as a fidelity optimization,
\begin{equation}
P_{\rm g}(Z|E)
=
\max_{\rho_{AB}\in\mathcal{S},\,\sigma_{AB}}
F^2\!\left(\rho_{AB},\,\mathcal{Z}(\sigma_{AB})\right),
\label{Coles01}
\end{equation}
where $F(\cdot,\cdot)$ denotes the Uhlmann fidelity and $\mathcal{Z}$ is the pinching map associated with Bob's $\mathbb{Z}$ measurement,
\begin{equation}
\mathcal{Z}(\sigma_{AB})=Z_0\sigma_{AB}Z_0+Z_1\sigma_{AB}Z_1,
\qquad
Z_j=\mathbb{I}_A\otimes|j\rangle_B\langle j|.
\end{equation}

The state $\rho_{AB}$ represents the (a priori unknown) bipartite state shared by Alice and Bob in the entanglement-based description of the protocol. In a realistic implementation $\rho_{AB}$ is not reconstructed tomographically; rather, it is only constrained through directly accessible statistics, such as the maximally mixed marginal on Alice's system [Eq.~(\ref{rhoAmixed})] together with the estimated error rates [Eq.~(\ref{subb1})]. Conceptually, this ``compatible-set'' perspective---optimizing a security functional over all quantum states consistent with a limited set of observed data---is reminiscent of self-testing, where one seeks to certify the underlying state and measurements from observed correlations alone, up to local isometries~\cite{vsupic2020self,Gigena2025}.
By contrast, the density matrix $\sigma_{AB}$ is an auxiliary optimization variable introduced by the fidelity-based representation of the guessing probability.


The constrained maximization in Eq.~(\ref{Coles01}) can be formulated as a semidefinite program (SDP), as detailed in the Appendix.
Solving the SDP yields $P_{\rm g}$ as a function of the QBER; for the symmetric case $\mathrm{QBER}^{(\mathbb{X})}=\mathrm{QBER}^{(\mathbb{Z})}=p$ the numerical solution is consistent with the analytic expression reported below.

The optimal solution returned by the SDP indicates that the maximizing $\rho_{AB}$ is Bell-diagonal,
\begin{align}
\rho_{AB}
&=p_0|\Phi^+\rangle\langle\Phi^+|
+p_1|\Phi^-\rangle\langle\Phi^-|\nonumber\\
&+p_2|\Psi^+\rangle\langle\Psi^+|
+p_3|\Psi^-\rangle\langle\Psi^-|,
\end{align}
with $p_0+p_1=(1-p)/2$ and $p_2+p_3=p/2$.
Furthermore, we find 
$p_2 - p_3 = p (1-2p)/2$
and
$p_0 - p_1 = ( 2 p^2 -3p+1)/2$.
If we assume that also $\sigma_{AB}$ is Bell-diagonal, then $\mathcal{Z}( \sigma_{AB} )$ is diagonal in the $\mathbb{Z}$ basis,
$\mathcal{Z}( \sigma_{AB} ) = \frac{1}{2}\text{diag}(s, 1-s, s, 1-s)$.
In this way the fidelity is only a function of the $s \in [0,1]$. 

With this ansatz, the fidelity reads
\begin{align}
\sqrt{2} \,
F \left( \rho, \mathcal{Z}( \sigma ) \right) & = \sqrt{(1-p) p (1-s)} 
+ p \sqrt{1 - s} \nonumber \\
& \phantom{=}~ 
     + (1-p)\sqrt{s} 
     +\sqrt{(1-p)p s} \, ,
\label{F2}
\end{align}
and by solving $ dF/ds = 0$ we find the stationary point $s=1-p$, yielding
\begin{equation}
\label{exact}
P_\text{g}(Z|E) = \frac12 + \sqrt{p (1-p)} \, ,
\end{equation}
which matches with the results obtained through numerical optimization.

\section{Finite-size estimates of the key rate}
\label{kratecost}

We now combine parameter estimation, error reconciliation, and privacy amplification to obtain explicit finite-size key-rate expressions for the three proof techniques compared in this work.
Because the QBER is estimated from a finite sample, the security proof must include statistical fluctuations and a corresponding parameter-estimation failure probability.

Upon $N$ transmitted signals, a fraction of the detected data is disclosed to estimate the QBER.
Denoting by $f\in(0,1)$ the fraction used for parameter estimation, the remaining fraction $(1-f)$ is available for key generation.
From $\eta N_{\rm err}=\eta f N$ disclosed samples, let $n_e$ be the number of observed errors, yielding the empirical estimate
\begin{equation}
\overline{\mathrm{QBER}}=\frac{n_e}{\eta N_{\rm err}}.
\end{equation}
To obtain a composable confidence interval, we introduce $\delta>0$ such that the event $\mathrm{QBER}>\overline{\mathrm{QBER}}+\delta$ occurs with probability at most $\epsilon_{\rm PE}$.
Following~\cite{tomamichel2012tight}, a sufficient choice is
\begin{equation}
\delta >
\sqrt{\frac{1}{\eta f N}\ln\!\left(\frac{1}{\epsilon_{\rm PE}}\right)}\, .
\label{err}
\end{equation}
In the remainder we conservatively replace $\mathrm{QBER}$ by $\overline{\mathrm{QBER}}+\delta$ in the key-rate expressions.

We quantify information reconciliation by the leakage $\ell_{\mathrm{leak}}$ (in bits) and allow for a nonzero error-correction failure probability $\epsilon_{\rm EC}$.
In the asymptotic limit, the optimal leakage is given by the binary entropy $h_2(\mathrm{QBER})$~\cite{cover1999elements}.
To account for practical inefficiencies, we write
\begin{equation}
\ell_{\mathrm{leak}}=\gamma\,\eta(1-f)N\,h_2(\overline{\mathrm{QBER}}+\delta),
\end{equation}
where $\gamma>1$ models the reconciliation inefficiency.

In the following, we denote by $\epsilon_{\rm h}$ the failure probability of privacy amplification and by $\epsilon_s$ the smoothing parameter where applicable.
For each technique, the resulting key is composably secure with an overall failure probability obtained by combining $\epsilon_{\rm PE}$, $\epsilon_{\rm EC}$, $\epsilon_{\rm h}$ (and $\epsilon_s$ when present).

\subsection{Finite-size min-entropy (FME)}

For collective attacks, the per-signal guessing probability factorizes across independent rounds.
Accounting for sifting and loss, the number of signals used for key generation is $n=\eta(1-f)N$.
Using Eq.~(\ref{exact}) together with the conservative replacement $p\mapsto \overline{\mathrm{QBER}}+\delta$, we obtain
\begin{align}
r^{\rm FME}_N
&\ge
\eta(1-f)\!\left(
\log_2\!\frac{1}{P_{\rm g}(\overline{\mathrm{QBER}}+\delta)}
-\gamma\,h_2(\overline{\mathrm{QBER}}+\delta)
\right)\nonumber\\&
+\frac{2}{N}\log_2\!\bigl(\sqrt{2}\,\epsilon_{\rm h}\bigr),
\label{est1}
\end{align}
where $P_{\rm g}(\cdot)$ is given by Eq.~(\ref{exact}).
The corresponding overall failure probability can be bounded by
\begin{equation}
\epsilon_{\rm FME}=\epsilon_{\rm h}+\epsilon_{\rm EC}+\epsilon_{\rm PE}.
\end{equation}
The dependence on the block size $N$ enters through the statistical term $\delta$ in Eq.~(\ref{err}) and through the additive privacy-amplification term of order $1/N$.

\subsection{Asymptotic equipartition property (AEP)}

The AEP relates the smooth conditional min-entropy to the conditional von Neumann entropy with a correction that scales as $1/\sqrt{N}$~\cite{tomamichel2012framework}.
Under collective attacks one has
\begin{equation}
\frac{1}{N}H_{\min}^{\epsilon_s}(Z^n|E^n)
\ge
H(Z|E)-\frac{\Delta(\epsilon_s)}{\sqrt{N}},
\end{equation}
with
\begin{equation}
\Delta(\epsilon_s)=4\log_2(2+\sqrt{2})\sqrt{\log_2\!\left(\frac{2}{\epsilon_s^2}\right)}\, .
\end{equation}
For BB84 with symmetric error rates, the conditional entropy evaluates to
\begin{equation}
H(Z|E)=1-h_2(\overline{\mathrm{QBER}}+\delta),
\end{equation}
yielding the finite-size key rate
\begin{align}
r^{\rm AEP}_N
&=
\eta(1-f)\left(1-(1+\gamma)\,h_2(\overline{\mathrm{QBER}}+\delta)\right)\nonumber\\&
-\sqrt{\frac{\eta(1-f)}{N}}\,\Delta(\epsilon_s)
+\frac{2}{N}\log_2\!\bigl(\sqrt{2}\,\epsilon_{\rm h}\bigr).
\label{rateaep}
\end{align}
The overall failure probability may be bounded as
\begin{equation}
\epsilon_{\rm AEP}=\epsilon_s+\epsilon_{\rm h}+\epsilon_{\rm EC}+\epsilon_{\rm PE}.
\label{epsilonAEP}
\end{equation}

\subsection{Entropic uncertainty relation (EUR)}

Entropic uncertainty relations provide a direct lower bound on $H_{\min}^{\epsilon_s}(Z^n|E^n)$ in terms of Bob's correlation with Alice in a complementary basis~\cite{tomamichel2011uncertainty,tomamichel2012tight}.
For BB84, the preparation quality is $q=1$, and one obtains the finite-size key-rate expression
\begin{align}
r^{\rm EUR}_N
&\ge
\eta(1-f)\left(1-(1+\gamma)\,h_2(\overline{\mathrm{QBER}}+\delta)\right)\nonumber\\&
+\frac{2}{N}\log_2\!\bigl(\sqrt{2}\,\epsilon_{\rm h}\bigr),
\label{rateunc}
\end{align}
with overall failure probability as in Eq.~(\ref{epsilonAEP}).
As $N\to\infty$, both the AEP- and EUR-based rates converge to the standard asymptotic expression.

\section{Results}
We now compare the finite-size key-rate estimates obtained from the FME, AEP, and EUR approaches.

\begin{figure}[t]
    \centering
    \includegraphics[width=1\linewidth]{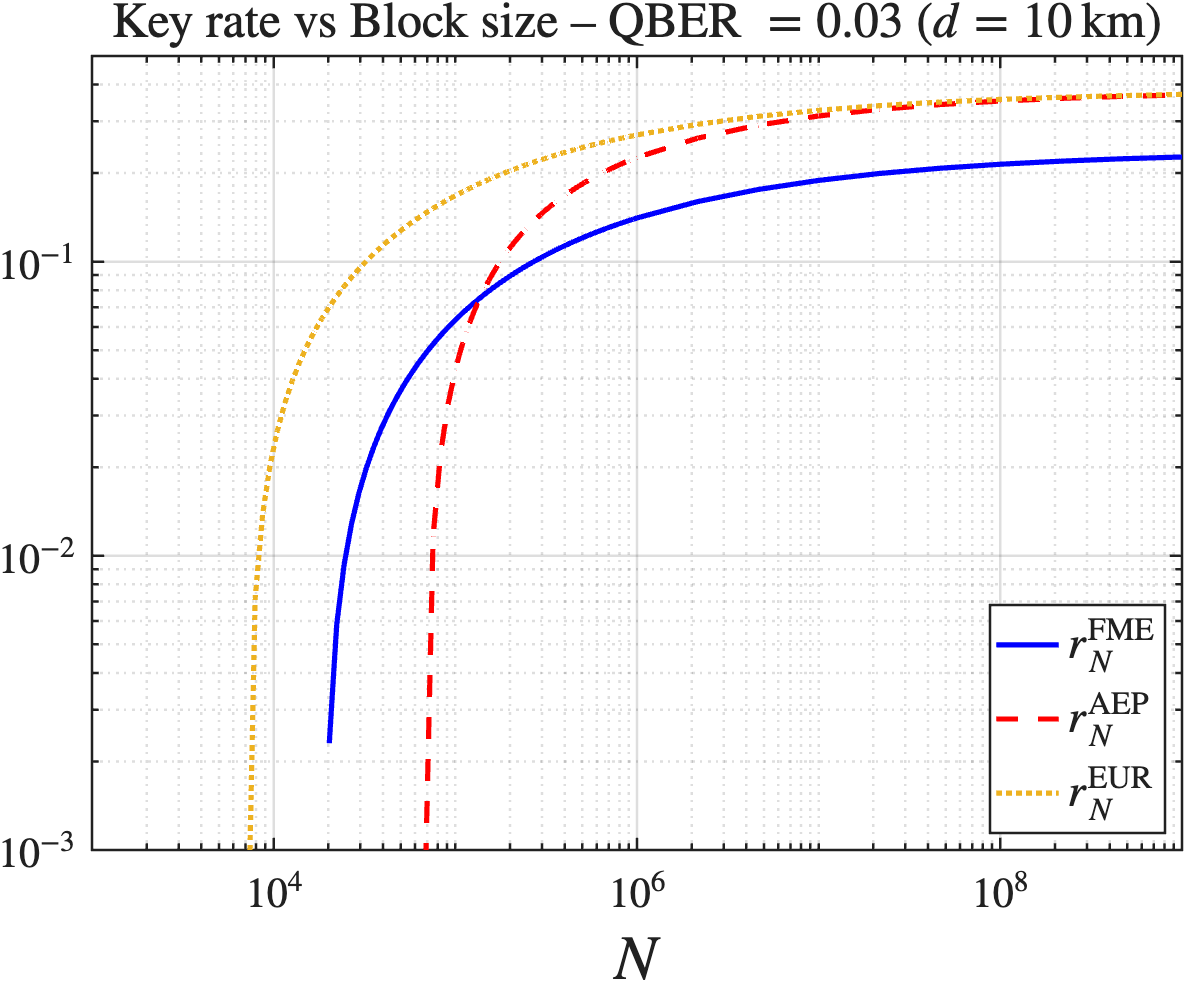}
    \caption{Key rates against block size of the protocol computed exploiting: 1) Guessing Probability, 2) AEP, 3) Entropic uncertainty relation. The rates are estimated for a distance of communication of $10 km$ and for $\text{QBER}=3\%$. The probabilities $\epsilon_{PE}$, $\epsilon_{EC}$, $\epsilon_{h}$, $\epsilon_{h}$ and $\epsilon_{s}$ are all set to $10^{-10}$. For details see~\cite{gioscaBB84}.}
    \label{kratep1}
\end{figure}

\begin{figure}[t]
    \centering
    \includegraphics[width=1\linewidth]{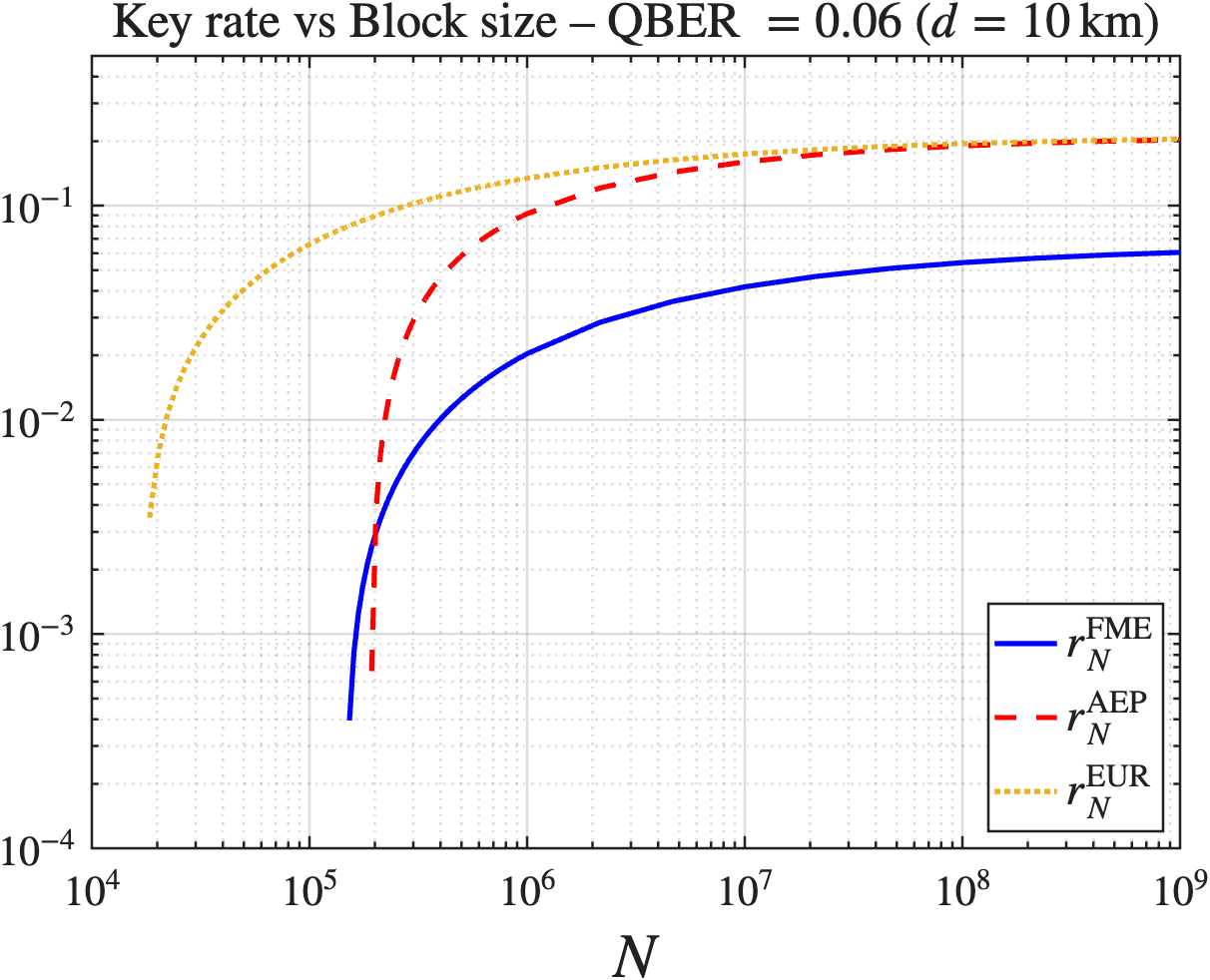}
    \caption{Key rates against block size of the protocol computed exploiting: 1) FME, 2) AEP, 3) EUR. The rates are estimated for a distance of communication of $10 km$ and for $\text{QBER}=6\%$. The probabilities $\epsilon_{PE}$, $\epsilon_{EC}$, $\epsilon_{h}$, $\epsilon_{h}$ and $\epsilon_{s}$ are all set to $10^{-10}$.}
    \label{kratep2}
\end{figure}



\begin{figure}[t]
    \centering
    \includegraphics[width=1\linewidth]{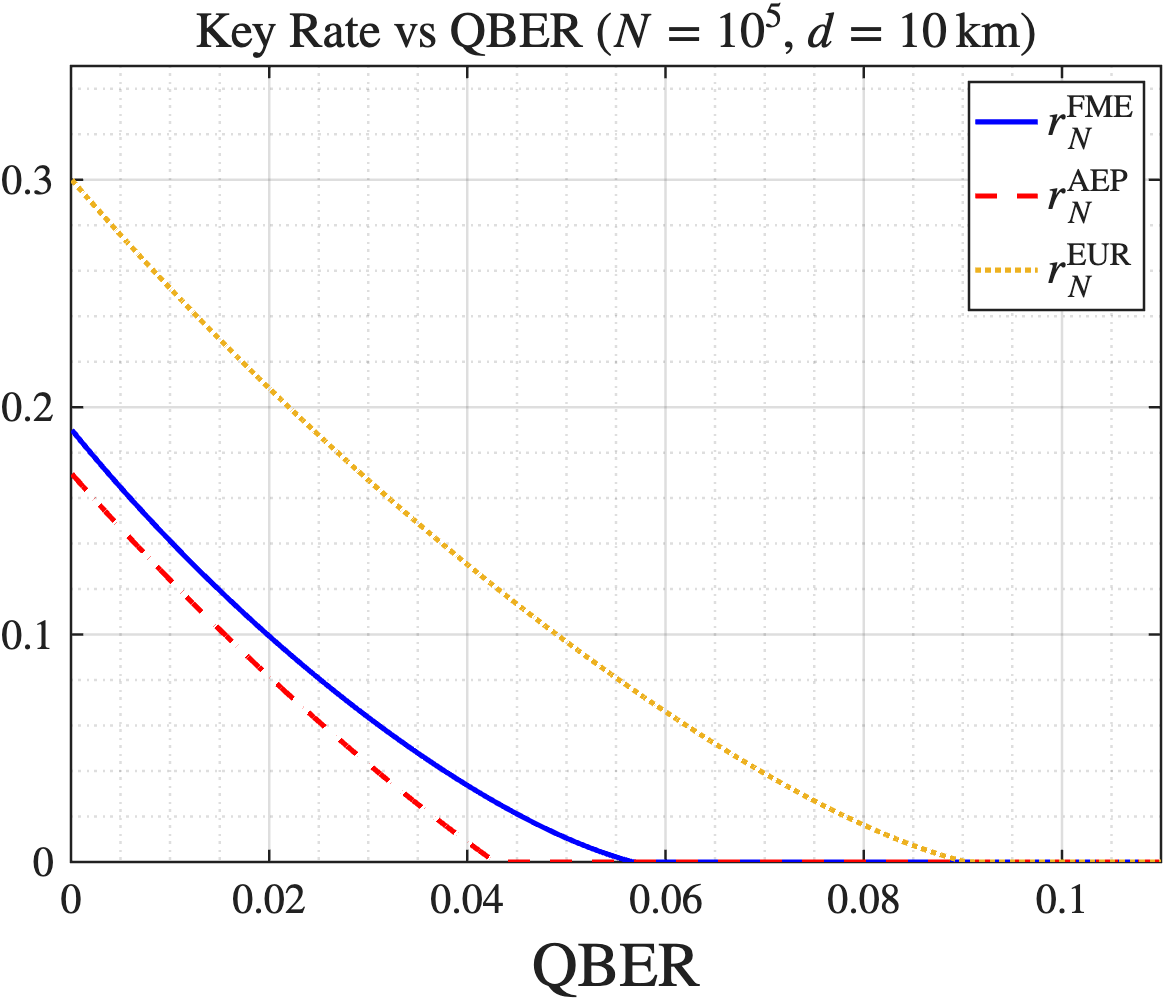}
    \caption{Key rates against QBER computed exploiting  1) Guessing Probability, 2) AEP, 3) Entropic uncertainty relation. The rates are estimated for a distance of communication of $10 km$ and for a block size $N=10^5$.}
    \label{kratepQB}
\end{figure}

\begin{figure}[t]
    \centering
    \includegraphics[width=1\linewidth]{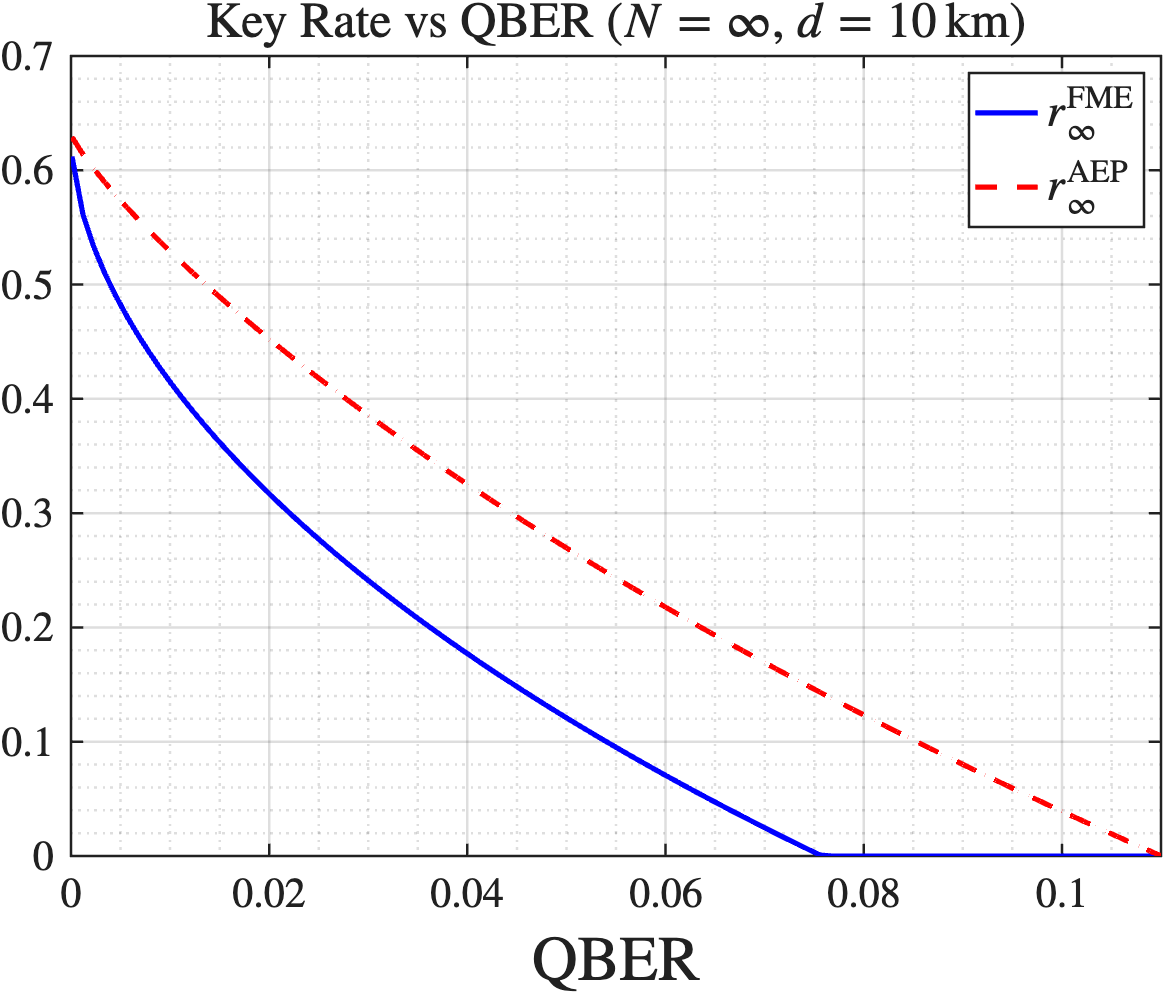}
    \caption{Key rates against $Qber$ in the asymptotic regime $N\longrightarrow\infty$ computed exploiting  1) Guessing Probability, 2) AEP. In the asymptotic limit AEP rate coincides with the one obtained through entropic uncertainty relation. The rates are estimated for a distance of communication of $10 km$.}
    \label{kratepQB2}
\end{figure}

In Fig.~\ref{kratep1} we report the secret-key rates as a function of the block size $N$ for a representative disturbance $\text{QBER}=3\%$ at a transmission distance $d=10\,\mathrm{km}$. We assume a fiber attenuation of $a=0.2\,\mathrm{dB/km}$, which is typical of terrestrial optical links~\cite{hiskett2006long}, and we numerically optimize the estimation fraction $f$ so as to balance the number of signals available for key extraction against the statistical uncertainty in parameter estimation. The EUR-based bound provides the highest rate over the whole range of $N$ considered, consistently with the fact that it does not incur the explicit $1/\sqrt{N}$ finite-size penalty that appears in the AEP correction term. The AEP-based bound approaches the same asymptotic expression as the EUR rate, but for moderate block sizes the additional $1/\sqrt{N}$ correction in Eq.~(\ref{rateaep}) can dominate and drive the certified rate to zero. By contrast, the FME estimate is obtained from a direct lower bound on the min-entropy via the guessing probability, and it does not contain such an explicit $1/\sqrt{N}$ term; as a result it may remain positive in parameter regimes where the AEP bound already collapses.

This comparison becomes more stringent at larger disturbances. In Fig.~\ref{kratep2} we repeat the analysis for $\text{QBER}=6\%$, keeping the same distance and attenuation model. For the conservative Chernoff/Hoeffding confidence level $\epsilon_{\mathrm{PE}}=10^{-10}$ adopted in the parameter-estimation step, the advantage of FME over AEP is quantitatively limited: in this higher-noise setting, FME exceeds AEP only in an intermediate window around $N\in[10^{5},2\times 10^{5}]$, while for smaller blocks both approaches return a vanishing rate. This sensitivity to the confidence interval highlights that the relative performance of different finite-key bounds depends not only on the entropic method itself, but also on how efficiently experimental statistics are converted into worst-case parameters.

Beyond the magnitude of the achievable key rate, it is also instructive to consider the maximum disturbance compatible with a non-vanishing key, which we use as a noise-tolerance figure of merit. Figure~\ref{kratepQB} shows the key rate as a function of $\text{QBER}$ in the finite-size regime at fixed block size $N=10^{5}$ and distance $d=10\,\mathrm{km}$. The EUR-based analysis tolerates the largest disturbance, yielding a positive rate up to $\text{QBER}\simeq 9\%$. In the same finite-size setting, the AEP-based bound drops to zero at $\text{QBER}\simeq 6\%$, whereas the FME-based estimate remains marginally positive at comparable disturbance levels, thus certifying key generation slightly beyond the point where the AEP rate already vanishes. Finally, in the asymptotic limit shown in Fig.~\ref{kratepQB2} there is no distinction between the AEP and EUR rates, as the AEP bound converges to the EUR expression when $N\to\infty$, and the corresponding threshold increases to $\text{QBER}\simeq 11\%$. The asymptotic FME bound is more conservative and predicts a smaller tolerable disturbance, with the rate vanishing around $\text{QBER}\simeq 8\%$.

\section{Conclusions}

We have compared three proof strategies that yield composable finite-key security bounds for quantum key distribution in the collective-attack model, focusing on entropic uncertainty relations (EUR), the asymptotic equipartition property (AEP), and a direct finite-size min-entropy approach (FME) based on the guessing probability. Using BB84 as a benchmark protocol, we derived finite-size key-rate estimates within a unified set of operational assumptions and highlighted the quantitative differences that emerge at realistic block lengths.

As expected, the EUR-based analysis provides the tightest key-rate estimates in the parameter regime relevant for discrete-variable QKD. The AEP-based approach is asymptotically optimal but introduces an explicit finite-size correction that decays only as $1/\sqrt{N}$, which can render the bound too pessimistic when the available block length is moderate. The FME approach, instead, evaluates the min-entropy directly in the finite-size setting and therefore avoids an explicit $1/\sqrt{N}$ penalty at the level of the entropy bound itself; in our BB84 case study this can allow one to certify a nonzero key in regimes where the AEP-based estimate already vanishes, although the resulting asymptotic performance remains below that of EUR.

Finite-key analyses rooted in min-entropy bounds have been explored previously. In particular, Bratzik \emph{et al.} showed that min-entropy techniques can yield nonzero BB84 key rates already for relatively small numbers of exchanged signals~\cite{bratzik2011min}, and Bunandar \emph{et al.} developed numerical finite-key methods that enable robust evaluations across a range of protocols and parameters~\cite{bunandar2020numerical}. Our contribution complements these works by providing a transparent benchmark between FME, AEP, and EUR within a consistent comparison framework and, for BB84 under symmetric error statistics, by identifying a closed-form expression for the optimal guessing probability. This analytic control clarifies when the direct min-entropy route is advantageous and helps isolate the role of parameter-estimation statistics in finite-size performance.
Going beyond min-entropy, Rényi entropies have been more recently explored for composable security in the finite-size regime~\cite{Dupuis10232924, Tan_arXiv2025, Cai2025} and in CV QKD protocols~\cite{PhysRevA.111.022610,staffieri2025finite}.

From a broader perspective, the value of the FME method may be most significant in settings where useful EUR-based bounds are not available or are not tight, as is often the case for continuous-variable protocols based on coherent states. In that context, security proofs against collective attacks typically rely on AEP-type arguments, and the resulting finite-size penalties can be prohibitive at practical block lengths. The present study suggests that direct finite-size min-entropy estimates may offer a complementary route to improve security analyses in such regimes.

The data that support the findings of this article are openly available \cite{gioscaBB84}.

\begin{acknowledgments}
This work has received support by the European Union's Horizon Europe research and innovation programme under the Project ``Quantum Secure Networks Partnership'' (QSNP, Grant Agreement No.~101114043)
and by INFN through the project ``QUANTUM''.
\end{acknowledgments}

\appendix

\appendix

\section{Semidefinite-program formulation for the guessing probability}
\label{app:sdp}

This Appendix summarizes the semidefinite-program (SDP) used to evaluate the guessing probability in Eq.~(\ref{Coles01}). The central ingredient is an SDP representation of the fidelity. For positive semidefinite operators $\rho$ and $\tau$ on a finite-dimensional Hilbert space $\mathcal{H}$, the fidelity
$F(\rho,\tau)=\mathrm{Tr}\sqrt{\sqrt{\rho}\,\tau\,\sqrt{\rho}}$
admits the characterization~\cite{watrous2018theory}
\begin{align}
F(\rho,\tau)
=
\max_{X\in\mathsf{L}(\mathcal{H})}
\Bigl\{
\mathrm{Re}\,\mathrm{Tr}(X)\;:\;
\begin{pmatrix}
\rho & X\\
X^\dagger & \tau
\end{pmatrix}\succeq 0
\Bigr\},
\label{eq:fidelity_sdp}
\end{align}
where $\succeq 0$ denotes positive semidefiniteness.
In our application, $\rho$ is the physical state $\rho_{AB}$ shared by Alice and Bob, while $\tau$ is the pinched operator $\mathcal{Z}(\sigma_{AB})$ associated with an auxiliary density operator $\sigma_{AB}$. The optimization in Eq.~(\ref{Coles01}) can therefore be implemented by introducing, in addition to $\rho_{AB}$ and $\sigma_{AB}$, an operator $X$ and enforcing the block positivity constraint in Eq.~(\ref{eq:fidelity_sdp}) with $\tau=\mathcal{Z}(\sigma_{AB})$. The pinching map $\mathcal{Z}$ is linear, so $\mathcal{Z}(\sigma_{AB})$ enters the SDP through linear matrix expressions.

The feasible set for $\rho_{AB}$ is defined by the protocol constraints and the observed statistics. Concretely, we enforce $\rho_{AB}\succeq 0$ and $\mathrm{Tr}(\rho_{AB})=1$, the marginal constraint $\mathrm{Tr}_B(\rho_{AB})=\mathbb{I}_A/2$, and the QBER constraints expressed as linear equalities of the form
\begin{align}
\mathrm{Tr}(\Pi^{(\mathbb{Z})}_{\mathrm{err}}\rho_{AB})=\text{QBER}^{(\mathbb{Z})},
\qquad
\mathrm{Tr}(\Pi^{(\mathbb{X})}_{\mathrm{err}}\rho_{AB})=\text{QBER}^{(\mathbb{X})},
\end{align}
with $\Pi^{(\mathbb{Z})}_{\mathrm{err}}=\ket{01}\!\bra{01}+\ket{10}\!\bra{10}$ and $\Pi^{(\mathbb{X})}_{\mathrm{err}}=\ket{\texttt{+-}}\!\bra{\texttt{+-}}+\ket{\texttt{-+}}\!\bra{\texttt{-+}}$. The auxiliary state satisfies $\sigma_{AB}\succeq 0$ and $\mathrm{Tr}(\sigma_{AB})=1$. When specializing to the symmetric case $\text{QBER}^{(\mathbb{X})}=\text{QBER}^{(\mathbb{Z})}=p$, the feasible set is further restricted accordingly.

For each value of $p$ we solve the resulting SDP to obtain the optimal fidelity $F(\rho_{AB},\mathcal{Z}(\sigma_{AB}))$, from which we compute $P_{\mathrm{g}}(Z|E)=F^2$ as in Eq.~(\ref{Coles01}). The numerical optimization was implemented in \textsc{Matlab} using CVX~\cite{MATLAB,cvx,gb08}.

\section{Finite-sample estimation of the QBER}
\label{app:qber}

In a finite-size implementation, the true error rate is not directly accessible and must be inferred from a random sample disclosed during parameter estimation. Let $m$ denote the number of detected signals used for this step; in our notation $m=\eta f N$. Writing $W_i\in\{0,1\}$ for the indicator variable of an error event in the $i$-th disclosed round, the empirical estimator is
\begin{align}
\overline{\text{QBER}}=\frac{1}{m}\sum_{i=1}^{m} W_i,
\end{align}
with $\mathbb{E}[W_i]=\text{QBER}$ under the i.i.d.\ assumption implicit in collective attacks.

A standard concentration inequality yields a confidence interval that holds except with probability $\epsilon_{\mathrm{PE}}$. For instance, Hoeffding's bound gives
\begin{align}
\Pr\!\left\{\text{QBER}\ge \overline{\text{QBER}}+\delta\right\}
\le \exp\!\left(-2m\delta^2\right),
\end{align}
so that choosing
\begin{align}
\delta=\sqrt{\frac{1}{2m}\ln\!\left(\frac{1}{\epsilon_{\mathrm{PE}}}\right)}
\label{eq:delta_hoeffding}
\end{align}
ensures $\text{QBER}\le \overline{\text{QBER}}+\delta$ with probability at least $1-\epsilon_{\mathrm{PE}}$. In the main text, this worst-case bound is used to convert the observed error frequency into a conservative QBER input for the entropic bounds.
\nocite{*}
\bibliographystyle{apsrev4-2}
\bibliography{bibliografia.bib}

\end{document}